\documentclass[10pt]{amsart}

\usepackage{amsmath,amssymb,dsfont}
\usepackage{graphicx}

\numberwithin{equation}{subsection}
\newfont{\bbd}{msbm10 scaled\magstep1}

\def\id{\hbox{{1}\kern-.25em\hbox{\rm l}}}
\def\one#1{#1^{\raise5pt\hbox{$\scriptstyle\!\!\!\!1$}}\,{}}
\def\two#1{#1^{\raise5pt\hbox{$\scriptstyle\!\!\!\!2$}}\,{}}
\def\half{\frac{1}{2}}

\def\comment#1{}

\def\?{(?)\marginpar{|?}}

\def\beq{\begin{equation}}
\def\eeq{\end{equation}}
\def\bea{\begin{eqnarray}}
\def\eea{\end{eqnarray}}
\def\bmat{\left(\begin{array}}
\def\emat{\end{array}\right)}

\newtheorem{rem}{Remark}
\newtheorem{prop}{Proposition}

\newtheorem{lemma}{Lemma}
\newcounter{subequation}[equation]
\makeatletter
\expandafter\let\expandafter
\reset@font\csname reset@font\endcsname
\def\subeqnarray{\arraycolsep1pt
    \def\@eqnnum\stepcounter##1{\stepcounter{subequation}%
        {\reset@font\rm(\theequation\alph{subequation})}}
\jot5mm     \eqnarray}

\makeatother

\newcounter{appendix}
\newcommand{\CH}{{\mathcal H}}
\newcommand{\CB}{{\mathcal B}}

\newcommand{\CZ}{{\mathcal Z}}

\newcommand{\CG}{{\mathcal G}}
\newcommand{\CM}{{\mathcal M}}

\newcommand{\CL}{{\mathcal L}}

\newcommand{\CX}{{\mathcal X}}
\newcommand{\CY}{{\mathcal Y}}

\def\be{\beta}
\def\al{\alpha}

\def\l{\lambda}
\def\la{\lambda}

\def\sig{\sigma}

\newcommand{\til}{\tilde}

\def\mat2#1#2#3#4{{\left(\begin{array}{cc}#1 & #2\\ #3 & #4
      \end{array}\right)}}

\def\mats2#1#2#3#4{{\left(\begin{array}{cc}#1 & #2\vspace{2truemm} \\ #3 & #4
\end{array}\right)}}

\def\endpf{\begin{flushright}$\square$\end{flushright}}

\begin{document}
\title[A rigid body dynamics derived from a class of 
  extended Gaudin models]%
  {A rigid body dynamics derived from a class of \\
  extended Gaudin models :
   an integrable discretization}

\author{Fabio Musso}
\address{Dipartimento di Fisica 'E Amaldi', Universit\'a degli Studi 'Roma Tre', 
and Istituto Nazionale di Fisica Nucleare, Sezione di Roma Tre, 
Via della Vasca Navale 84, Rome, Italy}
 \email{musso@fis.uniroma3.it}

\author{Matteo Petrera}
  \address{Dipartimento di Fisica 'E Amaldi', Universit\'a degli Studi 'Roma Tre',
and Istituto Nazionale di Fisica Nucleare, Sezione di Roma Tre, 
Via della Vasca Navale 84, Rome, Italy}
  \email{petrera@fis.uniroma3.it}

\author{Orlando Ragnisco}
\address{Dipartimento di Fisica 'E Amaldi', Universit\'a degli Studi 'Roma Tre', 
and Istituto Nazionale di Fisica Nucleare, Sezione di Roma Tre, 
Via della Vasca Navale 84, Rome, Italy}
 \email{ragnisco@fis.uniroma3.it}

\author{Giovanni Satta}
\address{Dipartimento di Fisica 'E Amaldi', Universit\'a degli Studi 'Roma Tre', 
and Istituto Nazionale di Fisica Nucleare, Sezione di Roma Tre, 
Via della Vasca Navale 84, Rome, Italy}
 \email{satta@fis.uniroma3.it}

\keywords{Gaudin models, B\"acklund transformations, spinning tops.}
%
%
\begin{abstract}
We consider a hierarchy of classical Liouville completely integrable models sharing the same
(linear) $r$--matrix structure obtained through an $N$--th jet--extension of $\mathfrak{su}(2)$
rational Gaudin models. The main goal of the present paper is the study of the integrable
model corresponding to $N=3$, since the case $N=2$ has been considered by the authors in separate 
papers, both in the one--body case (Lagrange top) and in the $n$--body one (Lagrange chain). 
We now obtain a rigid body associated with  a Lie--Poisson algebra which is an extension
of the Lie--Poisson structure for the two--field top, thus breaking its semidirect product structure.
In the second part of the paper we construct an 
integrable discretization of a suitable continuous Hamiltonian flow for the
system. The map is constructed following the theory of B\"acklund transformations
for finite--dimensional integrable systems developed by V.B. Kuznetsov and E.K. Sklyanin.
\end{abstract}

\maketitle
\section{Introduction} 

In \cite{MPR} we have considered  Liouville completely integrable Hamiltonian systems with $N$ degrees
of freedom obtained through the $N$--th jet--extension of Gaudin models.

Such procedure allows one to construct a hierarchy of integrable models sharing the same
(linear) $r$--matrix structure, whose first elements are, in the  $\mathfrak{su}(2)$ case 
\cite{MPR,KPR} :
\begin{itemize}
\item $N=1$: an Euler top associated with the Lie--Poisson algebra $\mathfrak{su}(2)$ 
\cite{Au,Su,RSTS}.
\item $N=2$: a Lagrange top associated with the Lie--Poisson algebra 
$\mathfrak{e}(3)=\mathfrak{su}(2) \oplus_s \mathbb{R}^3$ \cite{Au,Su,RSTS}.
\end{itemize}
Furthermore, a direct sum procedure
allows to build  long--range chains of $n$ interacting bodies, with rational, trigonometric and elliptic
$r$--matrices \cite{MPR, MPR2}.
For instance:
\begin{itemize}
\item $N=1$: a Gaudin model associated with $\bigoplus_{i=1}^n \mathfrak{su}_i(2) $ \cite{G1,G2}.
\item $N=2$: a Lagrange chain  associated with 
$ \bigoplus_{i=1}^n \mathfrak{e}_i(3)$ \cite{MPR,MPR2}.
\end{itemize}
Moreover this construction 
can be generalized to any  finite--dimensional simple Lie algebra 
instead of $\mathfrak{su}(2)$ \cite{MPR}.

The aim of the present paper is to investigate the one--body system corresponding to $N=3$, 
whose underlying algebra  still includes  $\mathfrak{su}(2)$ and $\mathbb{R}^3$ 
as  proper subalgebras but is no longer a semidirect sum of subalgebras.
This fact suggests one to see the $N=3$ case as a slightly generalized version of
the Lagrange top. The corresponding $n$-boby system will be considered in a separate paper.

We have here to mention that, up to our knowledge,
the system considered in this paper has been introduced,
in a different framework, by J.L. Thiffeault and P.J. Morrison in \cite{T1} and it
is called the twisted Lagrange top. They study this model in the spirit of the dynamical
systems theory, so that they do not use a Lax pair and an $r$--matrix approach, as we do
in the present work.
In \cite{T1} they obtain this new kind of
integrable top adding a cocycle to the Lie--Poisson structure for the two--field top \cite{RSTS},
thus breaking its semidirect product structure. We remark that
in \cite{T1} the so--called twisted top remains a mathematical construction without a physical
interpretation.

Later on, O. Vivolo, in \cite{Viv}, constructs a Lax matrix for such system, called
here generalized Lagrange top. The integrability is proven by direct inspection since
an $r$--matrix approach is not used, and the author focuses his attention
on the study of the spectral
curve of the system through the algebraic--geometry machinery. The main goal in  \cite{Viv}
is the proof that the generalized Lagrange top has monodromy, as well as the standard Lagrange top,
so that it does not admit global action--angle variables.

The main purpose of the present paper is the study of these generalized Lagrange tops
using the $r$--matrix structure inherited from Gaudin models \cite{MPR}. We obtain
complete integrability for a large hierarchy of integrable systems by providing
a Lax pair and a linear $r$--matrix algebra. In the second part of the paper we
obtain an integrable discretization of a suitable continuous Hamiltonian flow for the
system. The map is constructed following the theory of B\"acklund transformations
for finite--dimensional integrable systems developed by V.B.  Kuznetsov and E.K.  Sklyanin
in the papers \cite{V1,V2}.

We would like  here to remark that our approach
allows one to consider also a generic finite--dimensional simple Lie algebra 
instead of $\mathfrak{su}(2)$ and a natural generalization to a $n$--body system
with rational, trigonometric and elliptic dependences on the spectral parameter \cite{MPR}.

\section{Jet-extensions of $\mathfrak{su}(2)$ rational Gaudin models }

Let us  consider one--body rational Lax matrices of the following form:
\beq
\CL (\l) = \sum_{\al=1}^3 \sig^\al\,\left[ b^\al + \sum_{i=0}^{N-1}f_i(\la)\,y_i^\al \right],
\qquad f_0 (\la) \doteq \frac{1}{\l},
\label{lax}
\eeq
where $b^\al  \in  \mathbb{R}$, $\al=1,2,3$ and
$$
\sigma^1 \doteq  \half \left(\begin{array}{cc}
0 & {\rm{i}}  \\
{\rm{i}} & 0
\end{array}\right), \quad
\sigma^2 \doteq  \half \left(\begin{array}{cc}
0 & 1  \\
-1 & 0
\end{array}\right), \quad
\sigma^3 \doteq \half \left(\begin{array}{cc}
{\rm{i}} & 0  \\
0 & -{\rm{i}} 
\end{array}\right).
$$
The 
$3N$ coordinate functions $y_i^\al$, $\al=1,2,3$,  satisfy the following Lie--Poisson algebra:
\beq
\mathfrak{g}^{(N)}: \; \left\{ y^\al_i, y^\be_j \right\}= \left\{
\begin{array}{cc}
\epsilon^{\al \be}_{\quad \gamma} \,  y^\gamma_{i+j} & \quad i+j < N,\\
0 & \quad i+j \geq N.
\end{array} \right. 
\label{LP}
\eeq
Here  $\epsilon^{\al \be}_{\quad \gamma}$ is the skew--symmetric tensor 
with $\epsilon^{1 2}_{\quad 3}=1$. Let us notice that $N$ is exactly the
order of the jet--extension of the Lie--Poisson algebra $\mathfrak{su}(2)$ \cite{MPR}. It
coincides with the number of degrees of freedom of the system.

The functions $ f_i(\la)$ are chosen as
$$
f_i(\lambda)=\sum_{\{{\bf{q}}_i\} } \prod_{k=1}^i
\frac{c_k^{q_k}}{q_k!} \left( \frac{d}{d \lambda}
\right)^{|{\bf{q}}_i|} f_0 (\la), \qquad i=1,\dots,N-1, 
$$
where the $c_k$'s are arbitrary complex constants and
$$
\{{\bf{q}}_i\} \doteq \{ {\bf{q}} \in \mathbb{N}^i \ : \ q_1+2 \, q_2+\cdots+i\, 
q_i = i \}, \qquad |{\bf{q}}_i| \doteq \sum_{k=1}^i q_k, \qquad i=1,\dots,N-1.
$$

In \cite{MPR} we have shown that the following proposition holds.

\begin{prop} The Lax matrix (\ref{lax}) 
satisfies the linear $r$--matrix algebra
\beq
\left\{\CL(\l) \otimes \mathds{1}, \mathds{1} \otimes \CL(\mu) \right\} =
\left[r(\l - \mu), \CL(\l) \otimes \mathds{1} + \mathds{1} \otimes \CL(\mu) \right],\label{er}
\eeq
$$
r(\la)=\frac{1}{ \l} \sum_{\al=1}^3  \sig^{\al} \otimes \sig^{\al},
$$
whit respect to the Lie--Poisson algebra (\ref{LP}). Here $\mathds{1}$ is the $2 \times 2$ identity matrix.
\end{prop} 

The Lax matrix (\ref{lax}) can be written  in an equivalent form as a $2 \times 2$ matrix with elements
in the negative part of the loop--algebra $\mathfrak {g}^{(N)}[\lambda,\lambda^{-1}]$:
\beq
\CL (\l) = \frac{{\rm{i}}}{2}\left(\begin{array}{cc}
u(\l)  & v(\l)  \\
w(\l)  & -u(\l)
\end{array}\right),\label{yy}
\eeq
where
$$
u(\l) =  b^3 + \sum_{i=0}^{N-1}f_i(\la)\,y_i^3  , \quad
v(\l) = b^1 - {\rm{i}}\,  b^2 
+ \sum_{i=0}^{N-1} f_i(\la)\,\left( y_i^1 - {\rm{i}}\, y_i^2 \right)  , \quad
w(\l) = b^1 + {\rm{i}}\, b^2 
+ \sum_{i=0}^{N-1} f_i(\la)\,\left( y_i^1 + {\rm{i}}\, y_i^2 \right)  . 
$$
It is easy to see that
equation (\ref{er}) is equivalent to the following Lie-Poisson brackets for the rational
functions $u(\l)$, $v(\l)$, $w(\l)$:
\begin{eqnarray}
&&\{u(\l),u(\mu)\}=\{v(\l),u(\mu)\} =\{w(\l) ,w(\mu)\}=0, \nonumber \\
&&\{u(\l) , v(\mu) \}=\frac{{\rm{i}}}{\l - \mu} [v(\l) -v(\mu) ],\nonumber \\
&&\{u(\l) , w(\mu) \}=\frac{{\rm{i}}}{\l - \mu} [w(\l) -w(\mu) ],\nonumber  \\
&&\{v(\l) , w(\mu) \}=\frac{2\,{\rm{i}}}{\l - \mu} [u(\l) -u(\mu) ]. \nonumber
\end{eqnarray}

Recall that the correspondence
$$
(\xi^1,\xi^2,\xi^3)^T \in \mathbb{R}^3 \; \longmapsto \;
\xi = \half \left(\begin{array}{cc}
{\rm{i}} \, \xi^{3} &  {\rm{i}} \,\xi^{1}+ \xi^2 \\
{\rm{i}} \,\xi^{1}- \xi^2 & -\, {\rm{i}} \, \xi^{3}
\end{array}\right) \, \in \mathfrak{su}(2),
$$
 is an isomorphism between
$\mathfrak{su}(2)$ and the Lie algebra $(\mathbb{R}^3,[\cdot,\cdot] )$, where
the Lie bracket
$[\cdot,\cdot]$ is realized with the wedge product $\wedge$. Note that
$$
\langle \xi, \eta  \rangle = -2 \, {\rm{tr}} \, ( \xi \, \eta) = 2 \, {\rm{tr}} \, ( \eta \, \xi^*),
$$
where $\langle \cdot, \cdot \rangle$ is the scalar product in $\mathbb{R}^3$.

We now focus our attention on 
an interesting reduction of the Lax matrix (\ref{lax}). If we assume 
$c_1=-1$ and $c_k=0$ for $2 \leq k \leq N-1$ we readily get
\beq
\CL(\l) = \sum_{\al=1}^3 \sig^\al\,\left[ b^\al + \sum_{i=0}^{N-1}
\frac{y_i^\al }{\l^{i+1}} \right]. \label{lax2}
\eeq

Let us now fix the following notation: 
$$
{\bf{y}}_i \doteq (y_i^1,y_i^2,y_i^3)^T \in \mathbb{R}^3 \quad i=0,...,N-1, \qquad
{\bf{b}} \doteq (b^1,b^2,b^3)^T \in \mathbb{R}^3.
$$

The complete integrability of the hierarchy of systems obtained from the Lax matrices (\ref{lax2})
corresponding to different values of $N$ is established by the following statements.

\begin{lemma} The  characteristic curve $\Gamma^{(N)}: \, \det( \CL(\l) -\mu \, \mathds{1})=0$ 
is an hyperelliptic curve of the following form:
\beq
\Gamma^{(N)}: \, 4 \, \mu^2 + \langle{\bf{b}}, {\bf{b}} \rangle + H(\l) + C(\l)=0,\label{curve}
\eeq
where
\bea 
&& H(\l) = \sum_{i=1}^N \frac{1}{\l^i}
\left[  2 \, \langle{\bf{b}}, {\bf{y}}_{i-1} \rangle +
\sum_{k=1}^{i-2} \langle{\bf{y}}_{k}, {\bf{y}}_{i-k-2} \rangle
\right], \label{Hh}\\
&& C(\l) = \sum_{i=N+1}^{2N} \frac{1}{\l^i}
\sum_{k=i-N-1}^{N-1} \langle{\bf{y}}_{k}, {\bf{y}}_{i-k-2} \rangle.\label{C}
\eea
\end{lemma}

{\bf{Proof:}} A straightforward computation.

\endpf

\begin{prop} The curve (\ref{curve}) provides a  set of $2N$ Poisson--commuting integrals of motion
given by
$$
H_k= 2 \, \langle{\bf{b}}, {\bf{y}}_{k-1} \rangle + 
\sum_{i=0}^{k-2} \langle{\bf{y}}_i, {\bf{y}}_{k-i-2} \rangle, \qquad
C_k=
\sum_{i=k-1}^{N-1} \langle{\bf{y}}_i, {\bf{y}}_{N+k-i-2} \rangle, \qquad k=1,...,N,
$$
$$
\{ H_i, H_k \}=\{ C_i, H_k \}=\{ C_i, C_k \}=0, \qquad i,k=1,...,N.
$$
The integrals $H_k$, $k=1,...,N$ are first integrals of motion.
The integrals $C_k$, $k=1,...,N$ are linear combinations of the Casimir functions of the
Lie--Poisson algebra $\mathfrak{g}^{(N)}$,  namely
$$
\{ C_k, y^\beta_j \}=0 \quad \forall \, y^\beta_j \in \mathfrak{g}^{(N)}, \, 
\forall \, k=1,...,N.
$$
\end{prop}

{\bf{Proof:}} The quantities $H_k$ and $C_k$, $k=1,...,N$,  are immediately
obtained through the following formulae:
$$
H_k = {\rm{Res}}_{\l=0}\,  \l ^{k-1} H(\l), \qquad
C_k = {\rm{Res}}_{\l=0}\,  \l ^{N+k-1} C(\l), \qquad k=1,...,N,
$$
where $H(\l)$ and $C(\l)$ are given respectively in (\ref{Hh}) and (\ref{C}).
The function  $H(\l)$ provides $N$  Poisson--commuting first integrals of motion  thanks 
to the $r$--matrix 
structure (\ref{er}).
The fact that
$C(\l)$ is a generating function for the Casimirs of $\mathfrak{g}^{(N)}$ can be proven by a direct
computation:
\bea
\{ C_k, y^\beta_j \} &= & \sum_{\al=1}^3 \sum_{i=k-1}^{N-1}
\{ y_i^\al \,  y_{N+k-i-2}^\al , y^\beta_j \}=
\sum_{\al=1}^3 \sum_{i=k-1}^{N-1} \left[
 y_i^\al \{   y_{N+k-i-2}^\al , y^\beta_j \} + 
y_{N+k-i-2}^\al \{ y_{i}^\al , y^\beta_j \}
\right]= \nonumber \\
&=& 2 \sum_{\al=1}^3 \sum_{i=k-1}^{N-1} \epsilon^{\al \be}_{\quad \gamma} \, 
y_{i+j}^\gamma \, y_{N+k-i-2+j}^\al, \nonumber
\eea
for all $y^\beta_j \in \mathfrak{g}^{(N)}$, $k=1,...,N$.
Now, if $i+j \geq N$ then $\{ C_k, y^\beta_j \}$=0 thanks to (\ref{LP}). Let us consider
$i+j <N$:
$$
\{ C_k, y^\beta_j \} =
\sum_{\al=1}^3 \epsilon^{\al \be}_{\quad \gamma} \left[
\sum_{i=k-1}^{N-1} y_{i+j}^\gamma \, y_{N+k-i-2+j}^\al
+ \sum_{i'=k-1}^{N-1} y_{i'+j}^\al \, y_{N+k-i'-2+j}^\gamma
\right]=0,
$$
where $i'=N+k-i-2$. 
\endpf

\section{The 3-rd jet extension of $\mathfrak{su}(2)$ rational Gaudin models }

The Lax matrix 
\beq
\CL(\l)= \sum_{\alpha=1}^3
\sig^{\alpha} \left[b^\al + \frac{y_0^{\al}}{\l} 
-  \frac{c_1 \, y_1^{\al}}{\l^2 } + \frac{y_2^{\al}}{\l^2 } 
\left( \frac{c_1^2 }{\l} - c_2 \right)  \right], \label{kk}
\eeq 
is obtained considering $N=3$ in  (\ref{lax}).
Here $b^\al \in \mathbb{R}$, $\al=1,2,3$ plays the role 
of an  external field, 
taken as uniform  and constant (in time), $c_1,c_2$ are real arbitrary
constants and the  
9 coordinate functions $y_i^\al$, $\al=1,2,3$, $i=0,1,2$  satisfy the Lie--Poisson brackets
(\ref{LP}) with $N=3$ (i.e. $\mathfrak{g}^{(3)}$), namely:
\beq
\left\{    y_0^{\alpha},y_0^{\beta}   \right\}= \epsilon^{\al \be}_{\quad \gamma} \,y_0^{\gamma}, \qquad
\left\{    y_0^{\alpha},y_1^{\beta}   \right\}= \epsilon^{\al \be}_{\quad \gamma} \,y_1^{\gamma}, \qquad
\left\{    y_0^{\alpha},y_2^{\beta}   \right\}= \epsilon^{\al \be}_{\quad \gamma} \,y_2^{\gamma},
\label{algebra}
\eeq
$$
\left\{    y_1^{\alpha},y_1^{\beta}   \right\}=\epsilon^{\al \be}_{\quad \gamma} \, y_2^{\gamma}, \qquad
\left\{    y_1^{\alpha},y_2^{\beta}   \right\}= 0, \qquad
\left\{    y_2^{\al},y_2^{\be}   \right\}= 0. 
$$
Let us notice that
$$
\mathfrak{g}^{(3)} = \mathfrak{su}(2) \oplus_s \CG, \qquad {\rm{dim}}\, \CG = 6,
$$
where $\CG$, although including the abelian proper subalgebra $\mathbb{R}^3$
spanned by ${\bf{y}}_2$,
doesn't have a semidirect structure. In \cite{T1} it is shown that our algebra
$\mathfrak{g}^{(3)}$ can be obtained adding a cocycle to 
$\mathfrak{e}(3,2) = \mathfrak{su}(2) \oplus_s (\mathbb{R}^3 \oplus \mathbb{R}^3)$.
We recall that it is possible to use the Lie--Poisson algebra $\mathfrak{e}(3,2)$ to
describe a two--field top \cite{RSTS}.

The Lie--Poisson bracket between two functions
$f,g: \mathfrak{g}^{(3)} \rightarrow \mathbb{R}$ reads
\bea
\left\{f({\bf{y}}_0,{\bf{y}}_1,{\bf{y}}_2),g({\bf{y}}_0,{\bf{y}}_1,{\bf{y}}_2)   \right\}&=&
\langle {\bf{y}}_0, \nabla_{{\bf{y}}_0}\, f \wedge \nabla_{{\bf{y}}_0}\, g \rangle+
\langle {\bf{y}}_1, \nabla_{{\bf{y}}_0}\, f \wedge \nabla_{{\bf{y}}_1}\, g + 
\nabla_{{\bf{y}}_1}\, f \wedge \nabla_{{\bf{y}}_0}\, g \rangle+ \nonumber \\
&& + \, \langle {\bf{y}}_2, \nabla_{{\bf{y}}_0}\, f \wedge \nabla_{{\bf{y}}_2}\, g +
\nabla_{{\bf{y}}_2}\, f \wedge \nabla_{{\bf{y}}_0}\, g \rangle+
\langle {\bf{y}}_2, \nabla_{{\bf{y}}_1}\, f \wedge \nabla_{{\bf{y}}_1}\, g \rangle, \nonumber
\eea
where $\nabla$ is a gradient with respect to its subscript. The non semidirect structure
lies just in the last term of the above equation: this term is absent in the
$\mathfrak{e}(3,2)$ algebra.

The Lie--Poisson algebra $\mathfrak{g}^{(3)}$ has $3$ Casimir functions:
$$
C^{(1)} = \langle{\bf{y}}_0, {\bf{y}}_2 \rangle + \half \langle{\bf{y}}_1, {\bf{y}}_1  \rangle, \qquad
C^{(2)} = \langle{\bf{y}}_1, {\bf{y}}_2 \rangle, \qquad
C^{(3)} = \langle{\bf{y}}_2, {\bf{y}}_2 \rangle,
$$ 
which differ from the $\mathfrak{e}(3,2)$ Casimirs just for the presence of 
$\langle{\bf{y}}_0, {\bf{y}}_2 \rangle $ in $C^{(1)}$.

Specializing Lemma 1 to the $N=3$ case we immediately have the statement

\begin{prop} The characteristic curve $\Gamma^{(3)}: \, \det( \CL(\l) -\mu \, \mathds{1})=0$ provides
a set of 6 Poisson--commuting integrals of motion:
\beq
\Gamma^{(3)}: \, 4 \, \mu^2 + \langle{\bf{b}}, {\bf{b}} \rangle 
+ \frac{H_1}{\l}+\frac{H_2}{\l^2}+\frac{H_3}{\l^3}
+\frac{C_1}{\l^4}+\frac{C_2}{\l^5}+\frac{C_3}{\l^6}=0,\label{cu3}
\eeq
where
\bea
H_1 &=& 2 \, \langle{\bf{b}}, {\bf{y}}_0 \rangle , \nonumber \\
H_2 &=& \langle{\bf{y}}_0, {\bf{y}}_0 \rangle -
        2\,c_1 \, \langle{\bf{b}}, {\bf{y}}_1 + c_2 \,{\bf{y}}_2 \rangle, \nonumber \\
H_3 &=& 2\, c_1 \left(c_1 \,  \langle{\bf{b}}, {\bf{y}}_2 \rangle -
        \langle{\bf{y}}_0, {\bf{y}}_1 + c_2 \,{\bf{y}}_2 \rangle\right), \nonumber \\
C_1 &=&  c_1^2 \left( c_2^2 \, C^{(3)} + 2 \, c_2 \, C^{(2)} + 2\, C^{(1)}  \right), \nonumber \\
C_2 &=& -2 \, c_1^3 \left( c_2 \, C^{(3)} + C^{(2)} \right), \nonumber \\
C_3 &=& c_1^4 \,C^{(3)} . \nonumber 
\eea
\end{prop}

\section{A special reduction of the 3-rd jet extension}

Let us now consider the Lax matrix (\ref{lax2}) with $N=3$ and ${\bf{b}}=(0,0,b) \in \mathbb{R}^3$.
To simplify the notations we prefer
to rename the 
9 coordinate functions $y_i^\al$, $\al=1,2,3$, $i=0,1,2$ as
$y_0^\al \doteq y^\al$, $y_1^\al \doteq x^\al$, $y_2^\al \doteq z^\al$, $\al=1,2,3$.
Thus we obtain the following Lax matrix:
\beq
\CL(\l)= b \, \sigma^3 +\sum_{\alpha=1}^3
\sig^{\alpha} \left[ \frac{y^{\al}}{\l} 
+  \frac{x^{\al}}{\l^2 } + \frac{z^{\al}}{\l^3 }\right]. \label{kkk}
\eeq
Notice that (\ref{kkk}) is the extension to a third--order pole
(i.e. three degrees of freedom) of the Lagrange top Lax matrix \cite{KPR,RSTS}.
Thus, we may expect a generalization of the Lagrange system.

First, let us notice that the characteristic curve 
$\Gamma^{(3)}: \, \det( \CL(\l) -\mu \,\mathds{1})=0$ (\ref{cu3})
provides the following integrals of motion:
\beq
H_1 = 2 \, \langle{\bf{b}}, {\bf{y}} \rangle , \qquad
H_2 = \langle{\bf{y}}, {\bf{y}} \rangle +
        2\, \langle{\bf{b}}, {\bf{x}} \rangle, \qquad
H_3 = 2\, \left(  \langle{\bf{b}}, {\bf{z}} \rangle +
        \langle{\bf{y}}, {\bf{x}} \rangle\right), \label{we}
\eeq
\beq
C_1 =  2\, C^{(1)}= 2 \, \langle{\bf{y}}, {\bf{z}} \rangle + 
\langle{\bf{x}}, {\bf{x}}  \rangle, \qquad
C_2 = 2 \, C^{(2)}= 2\, \langle{\bf{x}}, {\bf{z}} \rangle, \qquad
C_3 = C^{(3)}= \langle{\bf{z}}, {\bf{z}} \rangle. \label{cas}
\eeq
As in the Lagrange case the third component of ${\bf{y}}$ 
and the euclidean norm of ${\bf{z}}$ are constants of the motion. Looking at 
the brackets (\ref{algebra}), and taking into account that ${\bf{y}}$ and 
${\bf{z}}$ span respectively
$\mathfrak{su}(2)$ and $\mathbb{R}^3$, we may interpret them as the
total angular momentum of the system and the vector pointing from
a fixed point (which we shall take as $(0,0,0) \in \mathbb{R}^3$)
to the centre of mass of an axially symmetric rigid body, namely
a Lagrange top.
Let us remark that ${\bf{y}}$ does not coincide with the
angular momentum of the top due to the presence of the vector ${\bf{x}}$.
We think of ${\bf{x}}$, whose norm in not constant,
as the position of the moving centre of mass of the global
system composed by the Lagrange top and a material point, whose position is described by
${\bf{x}}-{\bf{z}}$. Here we are assuming that both bodies have unitary masses. 
The link between 
 these two systems is given by integrals
$C_1$ and $C_2$.
If we think of a canonical realization of the Lie--Poisson algebra
$\mathfrak{g}^{(3)}$ (\ref{algebra}) in term of three canonical coordinates and their
conjugated momenta we can immediately argue that the vector ${\bf{x}}$
must depend on momenta, since
$\left\{    x^{\alpha},x^{\beta}   \right\}=\epsilon^{\al \be}_{\quad \gamma} \, z^{\gamma}$.
We come back to this point in more detail in section 5.

If we look at the first integral
$$
H_2 = \langle{\bf{y}}, {\bf{y}} \rangle + 2\, \langle{\bf{b}}, {\bf{x}} \rangle,
$$
we immediately see that it coincides with the physical Hamiltonian of the
Lagrange top, where  the vector
${\bf{y}}$ is the angular momentum of the spinning top and 
the vector ${\bf{x}}$ describes the motion
of the centre of mass of the the top on the surface $|{\bf{x}}|=c$, with $c$ constant.
In the case of the Lagrange top the equations of motion (in the rest frame) 
with respect to the Hamiltonian $H_2$
can be written in the 
following form:
$$
\left\{ \begin{array}{ll}
\dot {\bf{y}} = {\bf{b}} \wedge {\bf{x}}, \\
\dot {\bf{x}} = {\bf{y}} \wedge {\bf{x}},
\end{array}\right.
$$
which indicate that ${\bf{x}}$ rotates rigidly. In the following subsection we shall
derive the equations of motions for the system described by (\ref{kkk}) emphasizing
the main differences with the Lagrange case.

\subsection{A Lax representation}

First, it is useful to introduce the following complex generators:
$$
y^\pm = y^1 \pm {\rm{i}}\, y^2, \qquad x^\pm = x^1 \pm {\rm{i}}\, x^2, \qquad
z^\pm = z^1 \pm {\rm{i}}\, z^2.
$$
In term of $(y^3,y^\pm),(x^3,x^\pm),(z^3,z^\pm)$
the Lie--Poisson algebra $\mathfrak{g}^{(3)}$
reads
$$
\{y^{3},y^{\pm}\}= \mp\,  {\rm{i}} \,  y^{\pm}, \qquad
\{y^{+},y^{-}\}= -2 \, {\rm{i}} \, y^{3}, 
$$
$$
\{y^{3},x^{\pm}\}=\{x^{3},y^{\pm}\}=\mp \, {\rm{i}}\,  x^{\pm},\quad
\{y^{+},x^{-}\}=\{x^{+}, y^{-}\}=-2 \, {\rm{i}} \,x^{3}, \quad
\{y^{3},x^{3}\}=\{y^{+},x^{+}\}=\{y^{-},x^{-}\}=0,
$$
$$
\{y^{3},z^{\pm}\}=\{z^{3},y^{\pm}\}=\mp \, {\rm{i}}\,  z^{\pm},\quad
\{y^{+},z^{-}\}=\{z^{+}, y^{-}\}=-2 \, {\rm{i}} \,z^{3}, \quad
\{y^{3},z^{3}\}=\{y^{+},z^{+}\}=\{y^{-},z^{-}\}=0,
$$
$$
\{x^{3},x^{\pm}\}= \mp\,  {\rm{i}} \,  z^{\pm}, \qquad
\{x^{+},x^{-}\}= -2 \, {\rm{i}} \, z^{3}, 
$$
$$
\{x^{\alpha},z^{\beta}\}=\{z^{\alpha},z^{\beta}\}=0,  \qquad \al,\beta=\pm,3.
$$
The Lax matrix (\ref{kkk}) has the following simple form:
\begin{equation}
\CL(\l)= \CB 
+ \frac{\CY}{\l}+ \frac{\CX}{\l^2} + \frac{\CZ}{\l^3} ,
\end{equation}
where 
$$
\CB \doteq \frac{{\rm{i}}}{2}\left(\begin{array}{cc}
 b &  0 \\
 0 & -b
\end{array}\right), \quad 
\CY \doteq \frac{{\rm{i}}}{2}\left(\begin{array}{cc}
 y^{3} &  y^{-} \\
 y^{+} & - y^{3}
\end{array}\right), \quad
\CX \doteq \frac{{\rm{i}}}{2} \left(\begin{array}{cc}
 x^{3} & x^{-} \\
x^{+} & -x^{3}
\end{array}\right), \quad
\CZ \doteq \frac{{\rm{i}}}{2} \left(\begin{array}{cc}
 z^{3} & z^{-} \\
z^{+} & -z^{3}
\end{array}\right),
$$
are all $\mathfrak{su}(2)$ matrices.

We choose as a physical Hamiltonian of the system the first integral $H_2$:
\beq
\CH = \half H_2 = \half \langle{\bf{y}}, {\bf{y}} \rangle + \langle{\bf{b}}, {\bf{x}} \rangle=
-{\rm{Tr}} \, \left(\CY^2 +2\, \CB \, \CX  \right).
\label{H}
\eeq
We see that the kinetic term is given by the norm of the total angular momentum
(up to a factor $1/2$), while the potential energy is given by the projection
of the total centre of mass vector onto the external field.

In the case of the Hamiltonian (\ref{H}) the equations of motion are given by
\beq
\left\{ \begin{array}{ll}
\dot {\bf{y}} = {\bf{b}} \wedge {\bf{x}}, \\
\dot {\bf{x}} = {\bf{y}} \wedge {\bf{x}} + {\bf{b}} \wedge {\bf{z}}, \\
\dot {\bf{z}} = {\bf{y}} \wedge {\bf{z}},
\end{array}\right. \label{em}
\quad \Leftrightarrow \qquad
\left\{ \begin{array}{ll}
\dot \CY = [\CB, \CX], \\
\dot \CX = [\CY,\CX] + [\CB, \CZ] , \\
\dot \CZ = [\CY,\CZ].
\end{array}\right.
\eeq
We immediately see that the vector ${\bf{x}}$ does not rotate rigidly, though ${\bf{z}}$
still does. Obviously, since $|{\bf{x}}|$ is no longer preserved, the integral
$C_2 = 2 \, C^{(2)}= 2\, \langle{\bf{x}}, {\bf{z}} \rangle$ does not imply that the angle
between ${\bf{x}}$ and ${\bf{z}}$ is constant.
Using equations (\ref{em}) we obtain that the evolution equation for the vector pointing
from $(0,0,0) \in \mathbb{R}^3$ to the position of the material point is given by:
$$
\frac{d}{dt}({\bf{x}}-{\bf{z}})= {\bf{y}} \wedge ({\bf{x}}-{\bf{z}})+ {\bf{b}} \wedge {\bf{z}}
$$

We conclude this section giving a Lax representation for equations of motion 
(\ref{em}). The proof is straightforward.

\begin{prop}

The Lax representation for equations (\ref{em}) is given by
$$
\dot \CL (\l) \doteq \frac{d}{dt}\CL(\l) =\left[  \CL(\l), \CM (\l)  \right], \qquad
\CM (\l) = \frac{\CX}{\l} + \frac{\CZ}{\l^2}.
$$
\end{prop}

\section{A canonical realization of the reduced system}

As we have shown in the previous section, our model is a Hamiltonian system. Our aim is now to
find a coordinate transformation on the symplectic leaves that makes the system canonical. We
will use three Euler angles $\theta \in [0, 2 \pi)$, $\phi \in [0, 2 \pi)$ and 
$\psi \in [0, \pi)$ with their canonical conjugate momenta
$ p_\theta, p_\phi$ amd $p_\psi$. 

Our canonical description is restricted to the following symplectic leaf:
\beq
\mathcal{O} 
\doteq \left\{ ({\bf{y}},{\bf{x}}, {\bf{z}}) \in \mathfrak{g}^{(3)} |
\; C^{(1)} = 0,\; C^{(2)} = 0, \; C^{(3)} = 1 \right\}. \label{fo}
\eeq

\begin{prop} 

A canonical realization of the Lie--Poisson
algebra $\mathfrak{g}^{(3)}$ restricted to the symplectic leaf $\mathcal{O}$ (\ref{fo}) is given by:
\bea
&& {\bf{y}} = \left(\sin \phi \, p_\theta + \cot \theta \, \cos \phi \, p_\phi -
        \frac{\cos \phi}{\sin \theta} \, p_\psi,
         - \cos \phi \, p_\theta + \cot \theta \, \sin \phi \, p_\phi -
          \frac{\sin \phi}{\sin \theta} \, p_\psi, p_\phi \right), \nonumber \\
&& {\bf{x}} = \sqrt{2 \, p_\psi } \, \left(
\sin \psi \, \sin \phi - \cos \theta \, \cos \psi \, \cos \phi,
-\sin \psi \, \cos \phi - \cos \theta \, \cos \psi \, \sin \phi,
-\sin \theta \, \cos \psi \right), \nonumber \\
&& {\bf{z}} = \left(\sin \theta \, \cos \phi, \sin \theta \, \sin \phi,\cos \theta \right). \nonumber
\eea
\end{prop}

{\bf{Proof:}} It is well--known \cite{MR} that the canonical realization of an angular momentum
${\bf{y}}$ and of a constant vector ${\bf{z}}$ such that $|{\bf{z}}|=1$ in the euclidean
space is given by
$$
{\bf{y}} = \left(\sin \phi \, p_\theta + \cot \theta \, \cos \phi \, p_\phi -
        \frac{\cos \phi}{\sin \theta} \, p_\psi,
         - \cos \phi \, p_\theta + \cot \theta \, \sin \phi \, p_\phi -
          \frac{\sin \phi}{\sin \theta} \, p_\psi, p_\phi\right),
$$
$$
{\bf{z}} = (\sin \theta \, \cos \phi, \sin \theta \, \sin \phi,\cos \theta),
$$
where $q \doteq (\theta, \phi, \psi)$ are the standard Euler angles and 
$p \doteq (p_\theta, p_\phi, p_\psi)$ are their
canonical conjugated momenta. Let us recall that if $f(q,p)$
and $g(q,p)$ are two arbitrary functions
then
$$
\{f(q,p), g(q,p)  \}= \sum_{i=1}^3 \left[
\frac{\partial f(q,p)}{\partial q_i}\frac{\partial g(q,p)}{\partial p_i}-
\frac{\partial f(q,p)}{\partial p_i}\frac{\partial g(q,p)}{\partial q_i} \right].
$$
Requiring the Lie--Poisson brackets 
(\ref{algebra}) restricted to the symplectic leaf $\mathcal{O}$ (\ref{fo})
we easily obtain
$$
{\bf{x}} = \sqrt{2 \, p_\psi } \, (
\sin \psi \, \sin \phi - \cos \theta \, \cos \psi \, \cos \phi,
-\sin \psi \, \cos \phi - \cos \theta \, \cos \psi \, \sin \phi,
-\sin \theta \, \cos \psi).
$$
\endpf

As we mentioned in section 3 we have finally obtained the vector ${\bf{x}}$ described in term
of canonical coordinates $\theta,\phi,\psi$ and the conjugated momentum
$ p_\psi$. In particular, we have $|{\bf{x}}|^2 = 2 \, p_\psi > 0$.
As a consequence the Hamiltonian $\CH$ (\ref{H}) describes a non--holonomic
dynamics.

For the sake of completeness we write the first integrals of motion using the
above canonical description.
The physical Hamiltonian (\ref{H})  takes the following form:
$$
\CH = \frac{p_\theta^2}{2} +
\frac{p_\psi^2 + p_\phi^2 - 2\,p_\psi\, p_\phi \, \cos \theta}{2 \sin^2 \theta}-
b \, \sqrt{2 \, p_\psi } \sin \theta \cos \psi.
$$
We see that the variable $\phi$ is cyclic as in the Lagrange case, while $\psi$
explicitly enters in the potential term.
The remaining first integrals of motion are:
\bea
&& I_1 = \half H_2 = p_\phi, \nonumber \\
&& I_2 = \half H_3 = \sqrt{2 \, p_\psi } \left[  p_\theta \, \sin \psi + 
(p_\psi - p_\phi \, \cos \theta) \cot \theta \, \cos \psi -
p_\phi \, \sin \theta \, \cos \psi
\right] + b \, \cos \theta. \nonumber
\eea

\section{An integrable discretization through B\"acklund transformations}

The theory of integrable maps got a boost when Veselov developed a theory of Lagrange
correspondences \cite{Ves1,Ves2}. These maps are symplectic multi--valued transformations
which have enough integrals of motion, this definition being a proper analog of the classical
Liouville integrability. In the main examples, studied by him and later by other authors,
the integrable maps are constructed as time--discretizations of classical integrable
models, see, for instance, \cite{Ves1,Ves2,Su}.
Moreover these correspondences associate to a given
solution of an integrable system a new solution, a property reminescent of 
B\"acklund transformations (BTs) for soliton equations.

In this paper we apply the theory of BT for finite--dimensional
integrable systems, developed by V.B. Kuznetsov, E.K.  Sklyanin and P. Vanhaecke in  the relevant
papers \cite{V1,V2}. Following this approach we look at BTs as special Poisson maps. 
It is possible to find an exhaustive list of the features of these BTs in \cite{V1,V2}. Some of
these are the following ones:
\begin{enumerate}
\item a BT is a Poisson map that preserves the
same set of integrals of motion as does the continuous flow which it discretizes;
\item a BT is given by explicit formulae rather then implicit
equations;
\item although a BT is multi--valued, it leads to a single--valued map on any
level manifold of the integrals of motion;
\item when one searches for the simplest BT of an integrable system, then one finds
a one--dimensional family $\{\CB_\eta | \eta \in \mathbb{C}\}$ of them. The B\"acklund
parameter $\eta$  is canonically conjugate to $\mu$, i.e.  
$\mu= -\partial F / \partial \eta$ with $F_\eta$ generating function of 
$\{\CB_\eta | \eta \in \mathbb{C}\}$. Here $\mu$ is bound to $\eta$ by the
equation of an algebraic curve (dependent on the integrals), which is
exactly the characteristic curve that appears in the linearization of the integrable
system. This property is called spectrality of the BT;
\item the explicit nature of a BT makes it purely iterative, so that
it is very
well suited as symplectic integrator for the underlying model. Here the parameter
$\eta$ is an adjustable discrete time step. 
\end{enumerate}

\subsection{One--point BT}

First, let us consider the Lax matrix of our reduced system in the form
(\ref{yy}),
where the entries of $\CL (\l)$ are given by
$$
u(\l) =  b+ \frac{y^3}{\l}+\frac{x^3}{\l^2}+\frac{z^3}{\l^3}  , \qquad
v(\l) =  \frac{y^-}{\l}+\frac{x^-}{\l^2}+\frac{z^-}{\l^3}, \qquad
w(\l) =  \frac{y^+}{\l}+\frac{x^+}{\l^2}+\frac{z^+}{\l^3}. 
$$
A one--point BT can be defined
as the following similarity transform on the Lax matrix $\CL(\l)$:
$$
\CB_\eta: \;\CL(\l)  
\longmapsto \CM(\l;\eta) \,  \CL(\l) \, \CM^{-1}(\l,\eta) \qquad \forall ~\l \in \mathbb{C},\quad
\eta \in \mathbb{C},
$$
with some generically non-degenerate $2 \times2$ matrix $\CM(\l,\eta)$, simply because a BT should
preserve
the spectrum of $\CL(\l)$. The parameter $\eta$ is called B\"acklund parameter.
 
We use $~ \til {}$ -notations for the updated variables, so that
$$
\CB_\eta: \;\CL(\l)  
\longmapsto \;\til \CL (\l) = \frac{{\rm{i}}}{2}\left(\begin{array}{cc}
\til u(\l)  & \til v(\l)  \\
\til w(\l)  & -\til u(\l)
\end{array}\right),
$$
$$
\til u(\l) =  b+ \frac{\til y^3}{\l}+\frac{\til x^3}{\l^2}+\frac{\til z^3}{\l^3}  , \qquad
\til v(\l) =  \frac{\til y^-}{\l}+\frac{\til x^-}{\l^2}+\frac{\til z^-}{\l^3}, \qquad
\til w(\l) =  \frac{\til y^+}{\l}+\frac{\til x^+}{\l^2}+\frac{\til z^+}{\l^3}.
$$

We shall consider the similarity transformation between $\CL(\l)$ and $\til\CL(\l)$, namely
\begin{equation}
\CM(\l;\eta) \, \CL(\l)=\til \CL(\l) \, \CM(\l;\eta) \qquad \forall ~\l \in \mathbb{C}, \quad
\eta \in \mathbb{C},
\label{rr}
\end{equation}
with the following intertwining matrix \cite{KPR,HKR}:
\begin{equation}
\CM(\l;\eta)=\left(\begin{array}{cc}
\l-\eta +p \, q& p \\
q & 1
\end{array}\right),\qquad \det \CM(\l;\eta) =\l - \eta.
\label{ansatz}
\end{equation}
Note that the number of zeros of $\det \CM$ is the number of essential B\"acklund
parameters. Moreover the variables $p$ and $q$ are indeterminate dynamical variables.

Comparing the asymptotics in $\l\rightarrow\infty$ in both sides of (\ref{rr}) we readily get
\beq
\qquad p=\frac{y^-}{2\, b}, \qquad
q=\frac{\til y^+}{2\, b}, \qquad \til y^3 =y^3.  \label{hh}
\eeq
If we want an explicit map from $\CL(\l)$ to $\til \CL(\l)$
we must express $q$ in term of the old variables. To
solve this problem one can use
the spectrality of the BTs \cite{V1,V2}. Equation (\ref{rr}) defines a map $\mathcal{B}_{P}$
parametrized by the point
 $P=(\eta,\mu) \in \Gamma^{(3)}$. Notice that there are two points on
$\Gamma^{(3)}$, $P=(\eta,\mu)$ and $Q=(\eta,-\mu)$, corresponding to the same $\eta$ and sitting
one above the other because of the hyperelliptic involution:
\begin{equation}
(\eta,\mu) \in \Gamma^{(3)}: \qquad \det(\CL(\eta)
-\mu \, \mathds{1})=0. \nonumber
\end{equation}
This spectrality property give us the formula \cite{KPR,HKR}
\beq
q=\frac{u(\eta)-\mu}{v(\eta)}=-\frac{w(\eta)}{u(\eta)+\mu},
\label{hhh}
\eeq
where $\eta$ and $\mu$ are bounded by the algebraic curve
$$
-\mu^2 = \frac{1}{4} 
\left( \langle{\bf{b}}, {\bf{b}} \rangle + \frac{H_1}{\eta}+\frac{H_2}{\eta^2}+\frac{H_3}{\eta^3}
+\frac{C_1}{\eta^4}+\frac{C_2}{\eta^5}+\frac{C_3}{\eta^6}\right),
$$
and the integrals $H_i$ and $C_i$, $i=1,2,3$,  are given respectively in (\ref{we}) and (\ref{cas}). 
Now the equation (\ref{rr}) gives an integrable Poisson map from $\CL(\l)$ to $\tilde \CL(\l)$:
\bea
&& \til  u(\l) = 
\frac{(\l -\eta +2\,p \, q)[u(\l) -q \, v(\l)]+p \, w(\l)}{\l -\eta}, \nonumber \\
&& \til  v(\l) = 
\frac{(\l -\eta +2\,p\,q)^2 v(\l) -2\,p\,(\l -\eta +2 \,p\,q) u(\l) -p^2 w(\l)}
{\l -\eta}, \nonumber \\
&& \tilde  w(\l) = 
\frac{w(\l) +2\,q \, u(\l) -q^2 v(\l)}
{\l -\eta}. \nonumber
\eea
Collecting the negative powers of $\l$ the above formulae can be rewritten as an explicit map
$$
\CB_{\eta}:\,  ({\bf{y}},{\bf{x}}, {\bf{z}})\; \longmapsto \; (\til{\bf{y}},\til{\bf{x}}, \til{\bf{z}}), 
$$
given by
\beq
\begin{array}{lll}
\tilde y^3 &  =&  y^3,  \vspace{.2cm}\\ 
\tilde y^- &  =&  x^- + (p\, q-\eta)\, y^- -2\, p\, y^3, \vspace{.2cm}\\
\tilde y^+ &  =&  2\, q\,  b,\vspace{.2cm}\\
\tilde x^3 &  =&  x^3 + p\,y^+ -q\,x^- - q\, (pq\,- \eta)\, y^- +2\,p\,q\,y^3, \vspace{.2cm}\\
\tilde x^- & =&  (2\,p\,q- \eta)\, x^- -2\,p\,x\,^3 -p^2 \,y^+
               +p\,q\, (p\,q-\eta)\, y^--2\,p^2\, q\, y^3, \vspace{.2cm}\\
\tilde x^+ & =&  y^+ - \frac{q}{p}(p\,q-\eta)\, y^-+2\,q\,y^3, \vspace{.2cm}\\  
\tilde z^3 & =&  z^3 - q \, z^- +2 \, p \, q \, x^3 -p \, q^2\,  x^- +p\, x^+
               +\eta \,  q \, (p\, q -\eta)\,  y^- +p\,  \eta \,
               y^+ + 2\,  \eta\,  p\, q\,  y^3, \vspace{.2cm}\\
\tilde z^+ & =&  2\,q\,x^3 -q^2\,x^- +x^+ + \eta \,\frac{q}{p}(p\,q-\eta)\, y^- +\eta \,y^+ +
              2 \, \eta \,  q \, y^3, \vspace{.2cm}\\
\tilde z^- & =&  (2\,p\,q -\eta)\,  z^- -2\,p\,  z^3 -2\,p^2\,q\,x^3 +p^2\,q^2\, x^- -p^2\, x^+
             -\eta\, p\,q (p\,q -\eta) y^-\, -  \vspace{.2cm} \\
           &  &  -p^2\, \eta \,y^+ - 2\, \eta\, p^2\, q\, y^3. 
\end{array} \label{mp}
\eeq

The following statement shows how the one--point BT can be written in a symplectic form
through a generating function. We restrict our BT $\CB_\eta$ to a symplectic leaf of the Lie--Poisson
structure by fixing values of the Casimir functions $C^{(1)},C^{(2)},C^{(3)}$:
$$
\mathcal{O} 
\doteq \left\{ ({\bf{y}},{\bf{x}}, {\bf{z}}) \in
\mathfrak{g}^{(3)}|
\; C^{(1)} = \gamma_1,\; C^{(2)} = \gamma_2, \; C^{(3)} = 1 \right\}. \label{leaf}
$$
Let us fix the following notation:
$$
\Psi \doteq (y^3,x^3,z^3)^T, \qquad  \til \Psi \doteq (\til y^3, \til x^3, \til z^3)^T,
$$
$$
\chi^- \doteq (y^-,x^-,z^-)^T, \qquad \til \chi^+ \doteq (\til y^+,\til x^+,\til z^+)^T.
$$
\begin{prop} The one--point BT $\CB_\eta |_{\mathcal{O}}$ can be arranged 
 in the form:
\bea
&& \Psi_i = 
\sum_{j=1}^3 \{ \Psi_i, \chi^-_j \} \,  \nabla_{\chi^-}^j F_\eta (\chi^- | \til \chi^+) , \label{e1} \\
&& \til \Psi_i = 
\sum_{j=1}^3 \{ \til \chi^+_j, \til \Psi_i \}\, 
\nabla_{\til \chi^+}^j F_\eta (\chi^- | \til \chi^+), \label{e2}
\eea
$i=1,2,3$,  where
\bea
F_\eta (\chi^- | \til \chi^+) &=&
\frac{y^- \, \til y^+}{2\,b}  + k \left(\frac{y^-}{z^-}+ \frac{\til y^+}{\til z^+} \right)
-\frac{(1+ \eta \, \gamma_2)^2}{4 \, k \,  \eta^2}+
\half \left(\frac{\gamma_2^2}{4}-\gamma_1 \right) \ln \left( \frac{k+1}{k-1} \right)- \nonumber \\
&& 
- \frac{1}{2 \, k} \left[ \til z^+ \, x^- + z^- \, \til x^+  - \eta \, \til x^+ \, x^- 
+\frac{x^-}{z^-}\left(\frac{x^-}{z^-} + \frac{\eta}{2} x^- \, \til z^+ - \gamma_2 \right) 
+\frac{\til x^+}{\til z^+} \left(\frac{\til x^+}{\til z^+} + 
\frac{\eta}{2} \til x^+ \,  z^- - \gamma_2 \right)
\right], \label{fu}
\eea
with
$$
k^2 =1+ \eta \,  z^-\, \til z^+.
$$
\end{prop}

{\bf{Proof:}} The Casimir functions $ C^{(1)},C^{(2)},C^{(3)}$ do not change under the map:
$$
\CB_\eta: \; (C^{(1)},C^{(2)},C^{(3)}) \longmapsto (\til C^{(1)},\til C^{(2)},\til C^{(3)}) =
(C^{(1)},C^{(2)},C^{(3)}).
$$
The above invariance allows one to exclude 6 variables, expressing
$y^+,x^+,z^+$ and  $\til y^-,\til x^-,\til z^-$ in term of the
components of the vectors $\Psi, \chi^-$ and $\til \Psi, \til \chi^+$:
\bea
 y^+ &  =&  \frac{1}{z^-} \left[  2\, \gamma_1 -1 
 -2 \, y^3 \, z^3 - \frac{y^-}{z^-} (1-(z^3)^2) \right],  \nonumber\\ 
 x^+ &  =&  \frac{1}{z^-} \left[  2\, \gamma_2 -2 \, x^3 \, z^3 - \frac{x^-}{z^-} (1-(z^3)^2) \right], \nonumber\\
 z^+ &  =&  \frac{1-(z^3)^2}{z^-},\nonumber\\
\tilde y^- &  =&  \frac{1}{\til z^+} 
\left[  2\, \gamma_1 -1  -2 \, \til y^3 \, \til z^3 - \frac{\til y^+}{\til z^+} (1-( \til z^3)^2) \right], \nonumber\\
\tilde x^- & =&  \frac{1}{\til z^+} 
\left[  2\, \gamma_2 -2 \, \til x^3 \, \til z^3 - \frac{\til x^+}{\til z^+} (1-( \til z^3)^2) \right], \nonumber\\
\tilde z^- & =&  \frac{1-(\til z^3)^2}{\til z^+}.  \nonumber
\eea
With the help of (\ref{hh}) we can rewrite equations (\ref{mp}) of the map in the following form:
\bea
\Psi_1 =  y^3 &=& \frac{y^-\,y^+}{2\,b}-\frac{1}{2\,k^3}\left\{ \half \left[ \eta \left(z^-\,
\til x^+ + x^-\, \til z^+  \right)- z^-\,\til z^+ \right]^2 + \eta^2\,\left[ (z^-)^2
 \til z^+\,\til y^+ + (\til z^+)^2 z^-\, y^- \right] +\right. \nonumber \\
&& +  \left( 1 + \half \eta\, \gamma_2\right)\left(z^-\,\til x^+ + x^-\,\til z^+ \right) -
   \left( \gamma_2+ 2\,\eta\,\gamma_1\right) z^-\,\til z^+ -
   \eta\,\left( x^-\,\til x^+ + z^-\,\til y^+ + \til z^+\,y^- \right) + \nonumber \\
&& \left. + \frac{1}{4}\left( \gamma_2^2 - 4\,\gamma_1 \right) \right\} \nonumber , \\
\Psi_2= x^3 &=& \frac{1}{2\,k} \left[ \til y^+\, x^- + \eta\,\left( x^-\,\til z^+ + \til x^+\, z^- 
 \right) - \til z^+\,z^- + \gamma_2 \right] , \nonumber \\
\Psi_3= z^3 &=& \frac{\til y^+\, z^-}{2\,w}+ k , \nonumber \\
\til \Psi_1 = \til y^3 &=& y^3 , \nonumber \\
\til \Psi_2 = \til x^3 &=& \frac{1}{2\,k} \left[ \til x^+\, y^- + 
\eta\,\left( x^-\,\til z^+ + \til x^+\, z^- 
 \right) - \til z^+\,z^- + \gamma_2 \right] , \nonumber \\
\til \Psi_3 = \til z^3 &=& \frac{\til z^+\, y^-}{2\,w}+ k, \nonumber
\eea
where $k^2 =1+ \eta \,  z^-\, \til z^+$.
It is now easy to check that the function 
$F_\eta (\chi^- | \til \chi^+)$ (\ref{fu}) satisfies equations (\ref{e1}), (\ref{e2}).
\endpf

\begin{rem} {\rm{ It is possible to use the theory of canonical transformations to show that 
$\CB_\eta$ has the spectrality property.
The spectrality property of a BT means that the two coordinates 
$\eta$ and $\mu$ parametrizing the map are conjugated variables, namely
$$
\mu = -\frac{\partial F_\eta}{ \partial \eta}\,,
$$
where $F_\eta$ is the generating function of the BT.

In our case, using equations (\ref{hh}), (\ref{hhh}) and (\ref{fu}),  we obtain
$$
\mu=u(\eta)-\frac{\til y^+}{2\, b} \, v(\eta)= 
-\frac{\partial F_\eta (\chi^- | \til \chi^+)}{\partial \eta},
$$
so that the spectrality property holds}}.
\end{rem}

\begin{rem} {\rm{ We have here to remark that the one--parameter
map (\ref{mp}) is a complex transformation, namely it is not a physical BT for our system.
Following \cite{KPR} we shall construct, in the next subsection, 
a physical map using two B\"acklund parameters.}}
\end{rem}

\subsection{Two--point BT}

According to \cite{KPR,HKR}, we construct a composite map which is a product of the map
$\mathcal{B}_{P_1} \doteq \mathcal{B}_{(\eta_1,\mu_1)}$ and
$\mathcal{B}_{Q_2} \doteq \mathcal{B}_{(\eta_2,-\mu_2)}$:
\begin{equation}
\mathcal{B}_{P_1,Q_2}=\mathcal{B}_{Q_2} \circ \mathcal{B}_{P_1}:~
\CL(\l) \stackrel{\mathcal{B}_{P_1}} {\longmapsto} \, \til \CL(\l)
\stackrel{\mathcal{B}_{Q_2}} {\longmapsto} \; \breve{\CL}(\l). \nonumber
\end{equation}
The two maps are inverse to each other when $\eta_1=\eta_2$ and $\mu_1=\mu_2$.
This  two-point BT 
is defined by the following discrete--time Lax equation:
\beq
\CM(\l;\eta_1,\eta_2) \, \CL(\l)=\breve{\CL}(\l) \,\CM(\l;\eta_1,\eta_2) 
\qquad \forall \l\in \mathbb{C}, \qquad 
\eta_1,\eta_2 \in \mathbb{C},
\label{128}\end{equation}
where the matrix $\CM(\l;\eta_1,\eta_2)$ is \cite{KPR,HKR}
\beq
\CM(\l;\eta_1,\eta_2)=\begin{pmatrix}\l-\eta_1+s \, t
&t\cr -s^2\,t+(\eta_1-\eta_2)\, s&\l-\eta_2-s\,t\end{pmatrix},
\label{mm} 
\eeq
$$
\det \CM(\l;\eta_1,\eta_2) =(\l - \eta_1)(\l - \eta_2).
$$
The spectrality property with respect to two fixed points $(\eta_1,\mu_1)\in \Gamma^{(3)}$
and $(\eta_2,\mu_2)\in \Gamma^{(3)}$ give
\bea
s&=&\frac{u(\eta_1)-\mu_1}{v(\eta_1)}
=\frac{\breve u(\eta_2)-\mu_2}
{\breve v(\eta_2)},\label{x} \\
t&=&\frac{(\eta_1-\eta_2)(u(\eta_1)+\mu_1)(u(\eta_2)-\mu_2)}
{(u(\eta_1)+\mu_1)w(\eta_2)-(u(\eta_2)-\mu_2)w(\eta_1)}=
\frac{(\eta_1-\eta_2)(\breve u(\eta_1)-\mu_1)
(\breve u(\eta_2)+\mu_2)}
{(\breve u(\eta_2)+\mu_2)\breve w(\eta_1)-
(\breve u(\eta_1)-\mu_1)\breve w(\eta_2)}.\label{X}
\end{eqnarray}
Now we have two 
B\"acklund parameters $\eta_1,\eta_2 \in \mathbb{C}$. It is possible to obtain several
equivalent formulae \cite{KPR,HKR} for
the variables $s$ and $t$ since the points
$(\eta_1,\mu_1)$ and $(\eta_2,\mu_2)$ belong to the spectral curve $\Gamma^{(3)}$,
i.e. are bound
by the following relations
$$
- (2 \, \mu_j) ^2= u^2(\eta_j)+v(\eta_j)\,w(\eta_j)=
\breve u^2(\eta_j)+\breve v(\eta_j)\,
\breve w(\eta_j),
\qquad j=1,2.
$$

Together with (\ref{x}) and (\ref{X}), the 
 formula (\ref{128}) gives an explicit two--point Poisson integrable map from $\CL(\l)$ to
$\breve{\CL}(\l)$.
The map is parametrized by  the two points $P_1$ and
$Q_2$.

Obviously, when $\eta_1=\eta_2$ (and $\mu_1=\mu_2$) the map turns
into an identity map. As we have shown in \cite{KPR}
the two--point BT can be reduced to a real Poisson integrable map
if the following condition holds:
$$
\eta_1=\bar\eta_2 \doteq \eta= \mathfrak{R}(\eta) + {\rm{i}}\, \mathfrak{I}(\eta)
\in \mathbb{C}. \label{gg}
$$
Therefore,  the two--point map leads to a physical BT $\CB_\eta$
with two real parameters. 

Let us now introduce the following notation
$$
X \doteq  ({\bf{y}},{\bf{x}}, {\bf{z}})^T \in \mathbb{R}^9, \qquad 
\breve X \doteq  
(\breve{\bf{y}},\breve{\bf{x}}, \breve{\bf{z}})^T \in \mathbb{R}^9.
$$
A direct computation based on the similarity transform (\ref{128}) shows that

\begin{prop} The two--point BT 
$\CB_{P_1,Q_2}|_{\eta_1=\bar\eta_2} : \; X \longmapsto \breve X$ is given by
\beq
\breve  X = \Phi(s,t; \eta) \, X + X_0(s,t; \eta), \label{2p}
\eeq
with
$$
\Phi(s,t;\eta)= \left(\begin{array}{ccc}
\mathds{1}_{3 \times 3}  & \mathds{O}_{3\times 3} & \mathds{O}_{3\times 3}\\
A(s,t;\eta)  & \mathds{1}_{3 \times3} & \mathds{O}_{3\times 3}\\
B(s,t;\eta)  & A(s,t;\eta) & \mathds{1}_{3\times 3} 
\end{array}\right),
$$
where $A(s,t;\eta)$ and $B(s,t;\eta)$ are two $3 \times 3$ dynamical matrices 
depending on the B\"acklund parameter $\eta$ and the parameters $s,t$ 
and $X_0(s,t; \eta)$ is a dynamical vector. The matrices $\mathds{1}_{3 \times 3} $
and $\mathds{O}_{3 \times 3} $ are respectively the $3 \times 3$ identity matrix
and the $3 \times 3$ zero matrix.
\end{prop}

The explicit expressions of $X_0(s,t; \eta), A(s,t;\eta), B(s,t;\eta)$ are rather complicated
and they are given in Appendix 1.

\begin{rem}
{\rm{ Notice that, despite its matrix formulation, the map (\ref{2p}) is a non linear transformation}}.
\end{rem}

\section{Concluding remarks}

We have considered a hierarchy of classical Liouville completely integrable models sharing the same
(linear) $r$--matrix structure obtained through an $N$--th jet--extension of $\mathfrak{su}(2)$
rational Gaudin models. The general procedure of such extension is presented in \cite{MPR}.

We have fixed $N=3$ obtaining a rigid body associated to a Lie--Poisson algebra which is an extension
of the Lie--Poisson structure for the two--field top. We have here to recall that this
classical system has been introduced in \cite{T1} where 
it is called the twisted Lagrange top, and furtherly investigated in \cite{Viv}
in the algebraic--geometry setting.

The novelty of our approach is the introduction of an  $r$--matrix formulation for this system.
Its  knowledge enables us to easily find a Lax formulation for the equation of motion, as well as 
a canonical realization in terms of Euler angles. Finally, through  the approach developed by 
V.B. Kuznetsov, E.K. Sklyanin and P. Vanhaecke \cite{V1,V2}, we
find  explicit BTs for the system. In Appendix $2$ we present a numerical
simulation of the real reduction of the two--point BT. 

Let us remark that another feature of our approach is the  natural possibility of constructing
$n$--body integrable chains starting from each  $N$--th jet--extension, both considering any simple 
Lie algebra and rational, trigonometric and elliptic dependances on the spectral parameter.
 This was done for the $N=2$ case in \cite{MPRS}. The case  $N=3$ will be considered in a separate paper.


\section*{Appendix 1: Explicit expressions of $X_0(s,t; \eta), A(s,t;\eta), B(s,t;\eta)$}

As we have shown in section 6.2 the two--point BT can be written in the following form:
$$
\CB_\eta: \; X \longmapsto \breve X = \Phi(s,t; \eta) \, X + X_0(s,t; \eta),
$$
where
$$
\Phi(s,t;\eta)= \left(\begin{array}{ccc}
\mathds{1}_{3 \times 3}  & \mathds{O}_{3\times 3} & \mathds{O}_{3\times 3}\\
A(s,t;\eta)  & \mathds{1}_{3 \times3} & \mathds{O}_{3\times 3}\\
B(s,t;\eta)  & A(s,t;\eta) & \mathds{1}_{3\times 3} 
\end{array}\right),
$$
and $X_0(s,t; \eta)$ is a dynamical vector depending on  $\eta,s,t$  just as like the $3 \times 3$ matrices $A(s,t;\eta)$ and $B(s,t;\eta)$.

Here we recall that the dynamical variables $s$ and $t$ can be obtained using the spectrality property
of the BT and are given by formulas (\ref{x}) and (\ref{X}), namely:
$$
s=\frac{u(\eta_1)-\mu_1}{v(\eta_1)}, \qquad
t=\frac{(\eta_1-\eta_2)(u(\eta_1)+\mu_1)(u(\eta_2)-\mu_2)}
{(u(\eta_1)+\mu_1)w(\eta_2)-(u(\eta_2)-\mu_2)w(\eta_1)}.
$$
The explicit expressions of $X_0(s,t; \eta)$ is given by
\bea
&& [X_0(s,t; \eta)]_1 = b \, (s\, \al_1 -t ), \nonumber \\
&& [X_0(s,t; \eta)]_2 = -b \, (s\, \al_1 +t ), \nonumber \\
&& [X_0(s,t; \eta)]_3=0, \nonumber \\
&& [X_0(s,t; \eta)]_4= b\, \left[\frac{s\, \al_1}{2}(\al_2+\mathfrak{R}(\eta))+
                         t\,(\al_2- 2 \, \mathfrak{R}(\eta))\right], \nonumber \\
&& [X_0(s,t; \eta)]_5= {\rm{i}}\, 
                       b\, \left[ \frac{s\, \al_1}{2}(\al_2+\mathfrak{R}(\eta))-
                       t\,(\al_2- 2 \, \mathfrak{R}(\eta))\right], \nonumber \\
&& [X_0(s,t; \eta)]_6= 2 \, b\, t\, s\, \alpha_1, \nonumber \\
&& [X_0(s,t; \eta)]_7= -\frac{{\rm{i}}\, b}{4} 
\left[(\mathfrak{I}(\eta) +  \mathfrak{R}(\eta))^2 (t+\al_1)
+8 \, s\, t\, \mathfrak{R}(\eta)(t-\al_1)
\right], \nonumber \\
&& [X_0(s,t; \eta)]_8= -\frac{b}{4} 
\left[(\mathfrak{I}(\eta) +  \mathfrak{R}(\eta))^2 (t - \al_1)
+ 8 \, s\, t\, \mathfrak{R}(\eta)(t+\al_1)
\right], \nonumber \\
&& [X_0(s,t; \eta)]_9= 2\, [X_0(s,t; \eta)]_6 \, \mathfrak{R}(\eta), \nonumber 
\eea
where we have introduced the quantities 
\bea
&&\al_1= 2 \, \mathfrak{I}( \eta) - s\, t, \nonumber \\
&& \alpha_2 = \al_1 - s\, t, \nonumber 
\eea
and $\mathfrak{I}( \eta), \mathfrak{R}( \eta)$ denote respectively the imaginary part and the
real part of the B\"acklund parameter $\eta$.

The entries of the dynamical matrix $A(s,t;\eta)$ are given by:
\bea
&& [A(s,t;\eta)]_{ii} = 0  , \quad i=1,2,3 ,\nonumber \\
&& [A(s,t;\eta)]_{12} = {\rm{i}}\, \al_2 = - [A(s,t;\eta)]_{21}  , \nonumber \\
&& [A(s,t;\eta)]_{13} = s\, \al_1 - t= - [A(s,t;\eta)]_{31} , \nonumber \\
&& [A(s,t;\eta)]_{23} = -{\rm{i}} \, (s\, \al_1 + t )= - [A(s,t;\eta)]_{32}, \nonumber 
\eea 
so that $A(s,t;\eta)$ is a skew--symmetric matrix. The matrix $B(s,t;\eta)$ has
a more complicated form: its entries are
\bea
&& [B(s,t;\eta)]_{11}=\half (1 - s^2)(\al_1^2 -t^2), \nonumber \\
&& [B(s,t;\eta)]_{12}= 
-\frac{{\rm{i}}}{2}[(t^2 - s^2\, \al_1^2)- 2 \, \al_2 \,\mathfrak{R}( \eta) ], \nonumber \\
&& [B(s,t;\eta)]_{13}= \half s \, \al_1 \, (\al_2 + \mathfrak{R}( \eta)), \nonumber \\
&& [B(s,t;\eta)]_{21}=-\frac{{\rm{i}}}{2}
[(t^2 - s^2\, \al_1^2)+ 2 \, \al_2 \,\mathfrak{R}( \eta) ] , \nonumber \\
&& [B(s,t;\eta)]_{22}=\half (1 + s^2)(\al_1^2 -t^2) \nonumber \\
&& [B(s,t;\eta)]_{23}=- \frac{{\rm{i}}}{2}
 s \, \al_1 \, (\al_2 + \mathfrak{R}( \eta)), \nonumber \\
&& [B(s,t;\eta)]_{31}=\half \left[
t \, (\al_2 + 2\, \mathfrak{R}( \eta)) +  s \, \al_1 \, (\al_2 - 2\, \mathfrak{R}( \eta)) \right]
, \nonumber \\
&& [B(s,t;\eta)]_{32}=\half \left[
t \, (\al_2 + 2\, \mathfrak{R}( \eta)) -  s \, \al_1 \, (\al_2 - 2\, \mathfrak{R}( \eta)) \right]
, \nonumber \\
&& [B(s,t;\eta)]_{33}=4\, s\, t\,  \al_1. \nonumber 
\eea
\section*{Appendix 2: Numerics}

In this appendix we present a 3D plot corresponding to the real reduction of the two--point BT
(\ref{2p}). It is obtained using a MAPLE 8 program that is a slightly different
version of the MATLAB program developed by V.B. Kuznetsov in \cite{KPR}.

The input parameters are:
\begin{itemize}
\item 
the intensity of the external field, i.e. $b$;
\item 
the B\"acklund
parameter $\eta= \mathfrak{R} (\eta) + {\rm{i}}\, \mathfrak{I} (\eta)$. Here 
$\mathfrak{I} (\eta )$ is the time--step of the discretization;
\item the number of iteration of the map, $N$;
\item the initial values of the coordinate functions 
$y^1,y^2,y^3,x^1,x^2,x^3,z^1,z^2,z^3$.
\end{itemize}
The output is a 3D plot of $N+N$ consequent points $(x^1-z^1,x^2-z^2,x^3-z^3)$ and $(z^1,z^2,z^3)$.
We remark that the vector  $(x^1-z^1,x^2-z^2,x^3-z^3)$ describes the position of the material point, as 
explained in section 4, and the vector $(z^1,z^2,z^3)$ is the position of the centre of mass of the 
spinning top.
As expected, the points $(z^1,z^2,z^3)$ lie on the sphere 
$C^{(3)}=\langle \bf{z},\bf{z} \rangle =$\,constant,
of some radius defined by the initial data.

\begin{figure}[h!]

\begin{center}
\includegraphics[height=8cm]{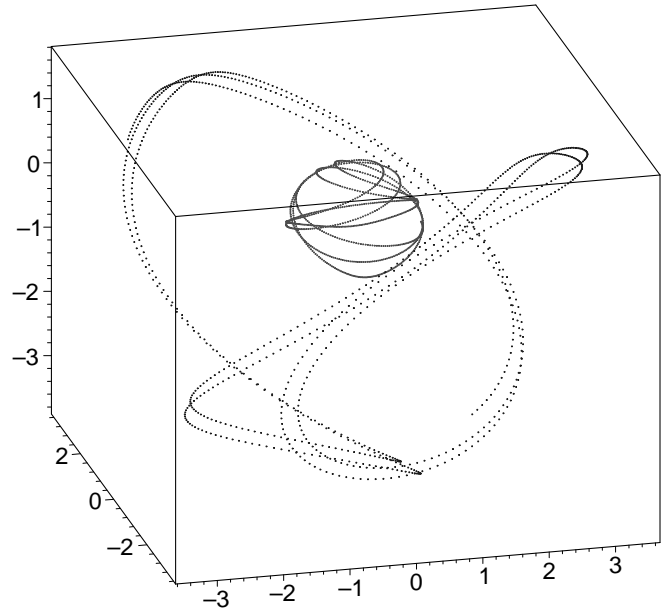}
\end{center}
\caption{$\scriptstyle 
(b;\mathfrak{R} (\eta),\mathfrak{I} (\eta);N; y^1,y^2,y^3,x^1,x^2,x^3,z^1,z^2,z^3)=
(1;5,0.1;1000;
-2.4,-0.6,-1.2,-2.19,0.89,1.34,1,0,0)$}

\end{figure}

\begin{figure}[h!]

\begin{center}
\includegraphics[height=8cm]{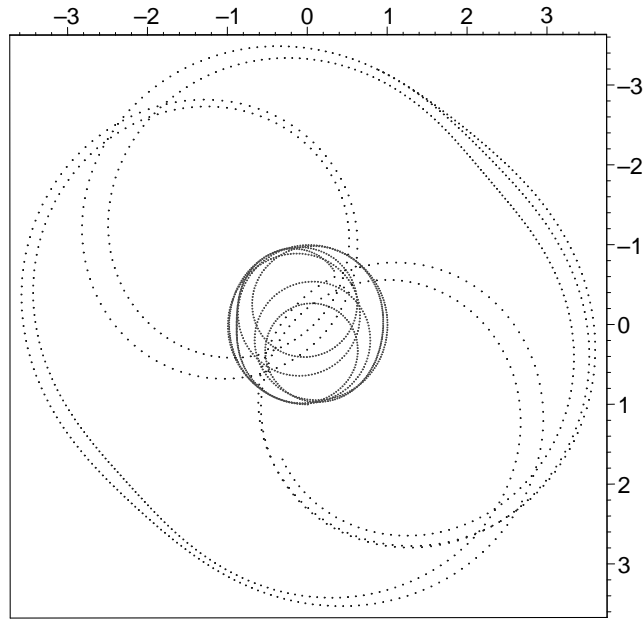}
\end{center}
\caption{Projection of Figure 1 on the $x-y$ plane}

\end{figure}

\newpage


\end{document}